\documentclass[10pt,a4paper,english]{article}

\input oejv.sty

\usepackage{tablefootnote}
\usepackage{float}

\usepackage{graphicx} 
\usepackage[graphicx]{realboxes}

\usepackage{subfig}

\usepackage{etoolbox}
\apptocmd{\sloppy}{\hbadness 10000\relax}{}{}

\begin{document}

\OEJVhead{Jan 2023}
\OEJVtitle{CANDIDATE TRIPLE-STAR SYSTEM WITH ELLIPSOIDAL COMPONENTS}
\OEJVtitle{DETECTED IN VULPECULA THROUGH TESS PHOTOMETRY}
\OEJVauth{G. Conzo$^1$; M. Moriconi$^1$; I. Peretto$^2$}

\OEJVinst{Gruppo Astrofili Palidoro, Fiumicino, Italy {\tt \href{mailto:ricerca@astrofilipalidoro.it}{ricerca@astrofilipalidoro.it}}}
\OEJVinst{MarSEC Marana Space Explorer Center, Crespadoro, Italy {\tt \href{mailto:ricerca@marsec.org}{ricerca@marsec.org}}}

\OEJVabstract{\normalsize{\textit{\textit{MaGiV-1}} is a candidate triple ellipsoidal star system in Vulpecula at coordinates RA(J2000) $19:52:19.13$ DEC(J2000) $+23:29:59.7$ classified as ELL+ELL, number 2344411 in AAVSO VSX database.
Through photometry from the TESS Space Telescope, two significant periods describing the orbital times of the components were identified using the Fourier transform. The analysis led to determining $P_{A-BC} = (4.269 \pm 0.213)d$ the orbital period of \textit{A-BC} pair, the primary component with the secondary component described by another pair, and $P_{BC} = (0.610 \pm 0.031)d$ the orbital period of \textit{B-C} pair, the inner ellipsoidal system. However, it cannot be completely ruled out that the shorter period can be explained by pulsations of one of the two components (e.g. by the GDOR type).
}
}

\begintext

\section{Introduction}\label{secintro}

\begin{sloppypar}Rotating ellipsoidal variable star systems have two stars that brightness variation shows up in photometric analysis from mutual deformation of the components by gravitational interaction. Therefore it is not generated from eclipses, but it is generated by cross-sectional area changes of the stars and it is represented by rather homogeneous light curves with an almost sinusoidal pattern and its variations are less than a tenth of a magnitude. It is very difficult to classify this type of variable star because the light curve profile does not contain the amount of information that light plots of eclipse binaries do \citep{1985ApJ...295..143M}.

This paper is about a very complex variability closely related to the ELL type (Ellipsoidal Variable Star), as it is a candidate system consisting of three ellipsoidal components describing a variation of the classical ellipsoidal binary star system. \textit{\textit{MaGiV-1}} has been defined as a ``double rotating ellipsoidal variable'' (ELL+ELL) star, a triple hierarchical type system, therefore, three stars orbiting each other simultaneously, and specifically two of these objects form a narrow binary, called inner binary, then the third companion is orbiting at a distance that far exceeds the separation length of the inner binary \citep{2017ApJ...834..200M}. Again, the brightness variation is represented by a light curve with variability less than a tenth of a magnitude, but, from the Fourier analysis for finding the period, two significant values emerge instead of one, thus returning two distinct light curves (one child of each other). Both resulting periods represent the orbital period for each pair, just as in binary ellipsoidal variables \citep{2011AcA....61..247J}.

\section{Photometric observations}

\begin{figure}[ht]
\centering
\includegraphics[width=14cm]{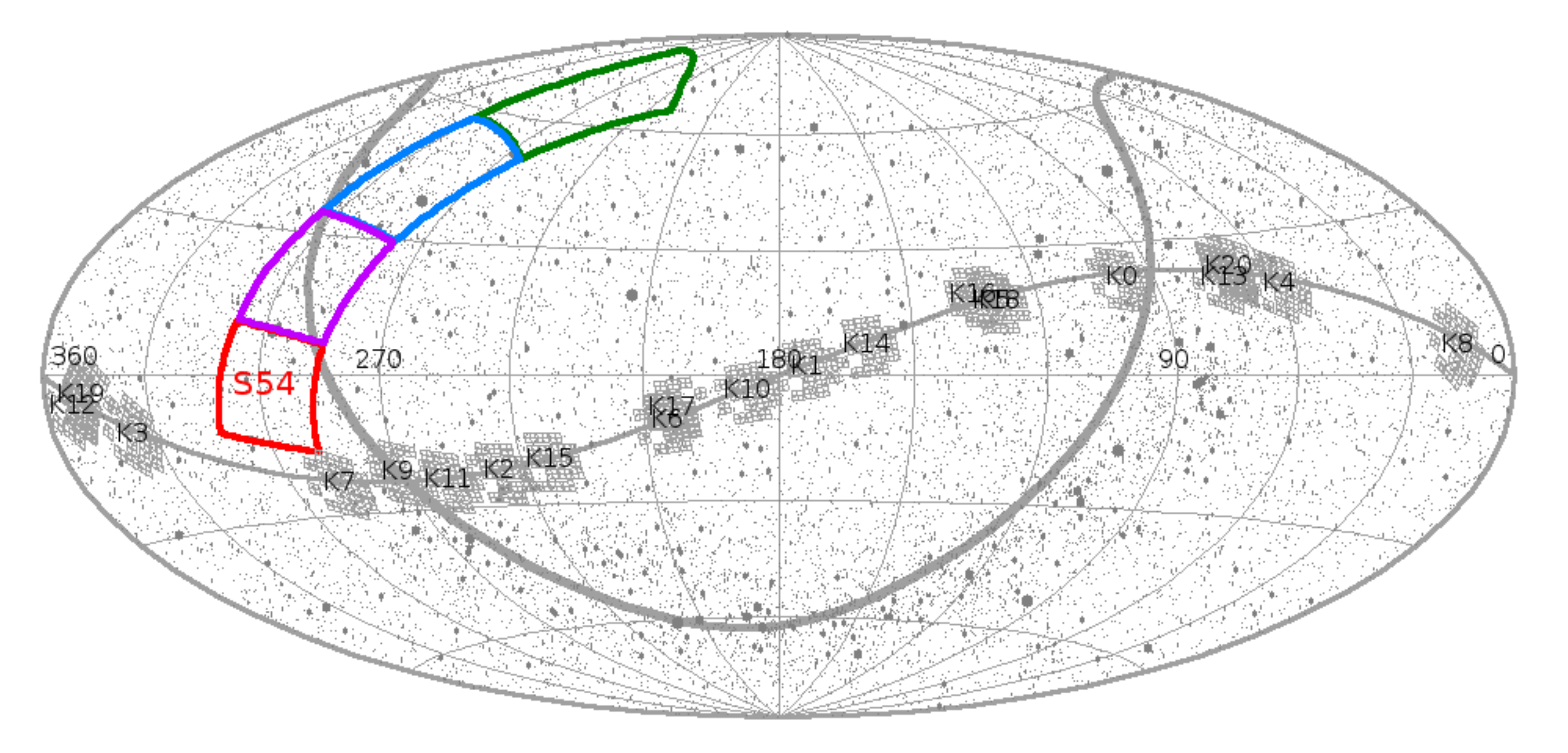}
\caption{Map of TESS camera fields-of-view in celestial coordinates \citep{2014SPIE.9143E..20R}. Sector 54.}
\label{fig:tesssec54}
\end{figure}

For detection of \textit{\textit{MaGiV-1}} system, TESS photometry \citep{RickerGR(2014)} was used, showing millesimal brightness changes. TESS observes the sky in sectors measuring $24^{\circ} \times 96^{\circ}$.  Each sector is observed for two orbits of the satellite around the Earth, or about 27 days on average. In our case, sector 41 and sector 54 were used, then photometrc observations derive from cameras as shown in Figure \ref{fig:tesssec41}  and Figure \ref{fig:tesssec54}:

\begin{figure}[ht]
\centering
\includegraphics[width=14cm]{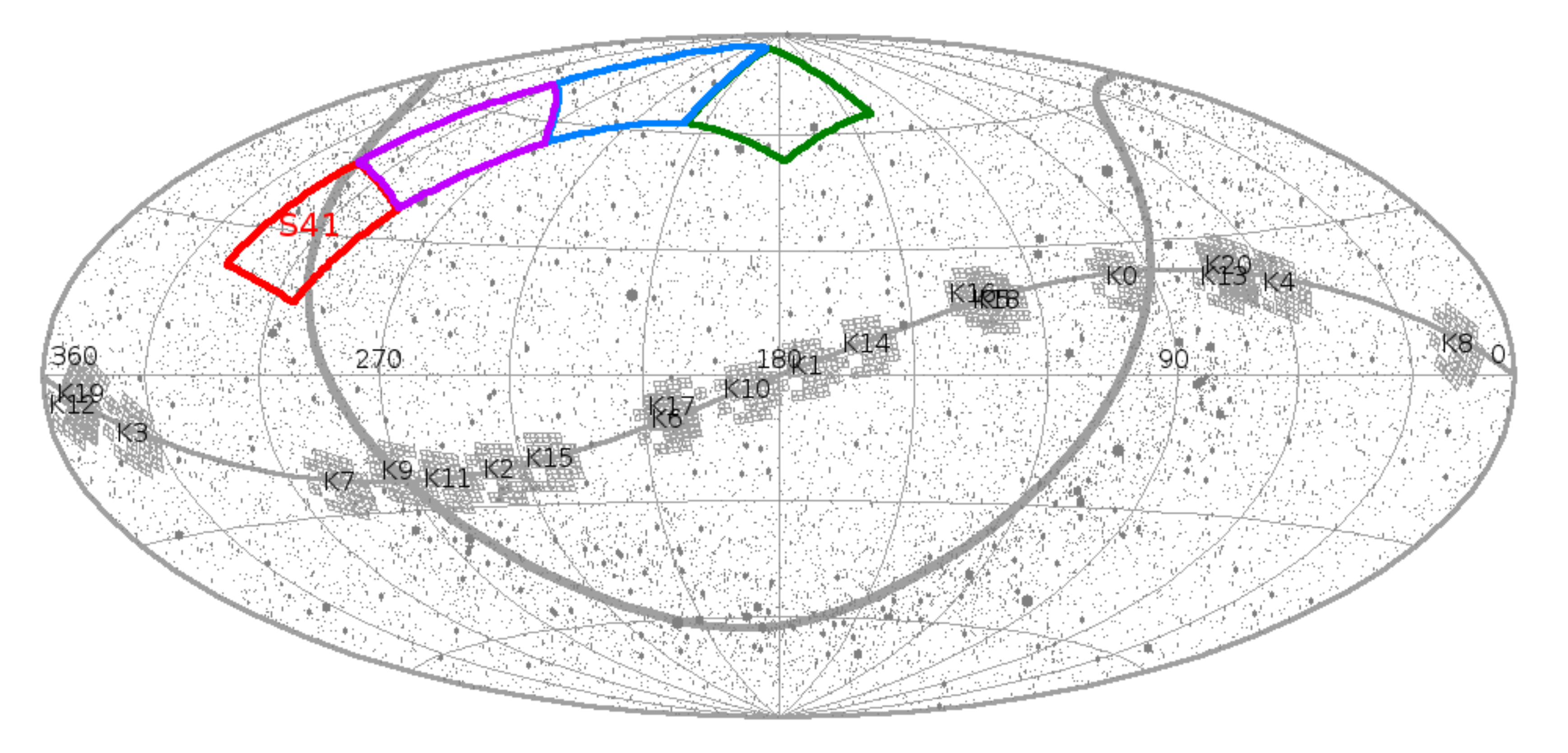}
\caption{Map of TESS camera fields-of-view in celestial coordinates \citep{2014SPIE.9143E..20R}. Sector 41.}
\label{fig:tesssec41}
\end{figure}

\newpage

\begin{figure}[ht]
\begin{minipage}[b]{7.5cm}
\centering
\includegraphics[width=7.5cm]{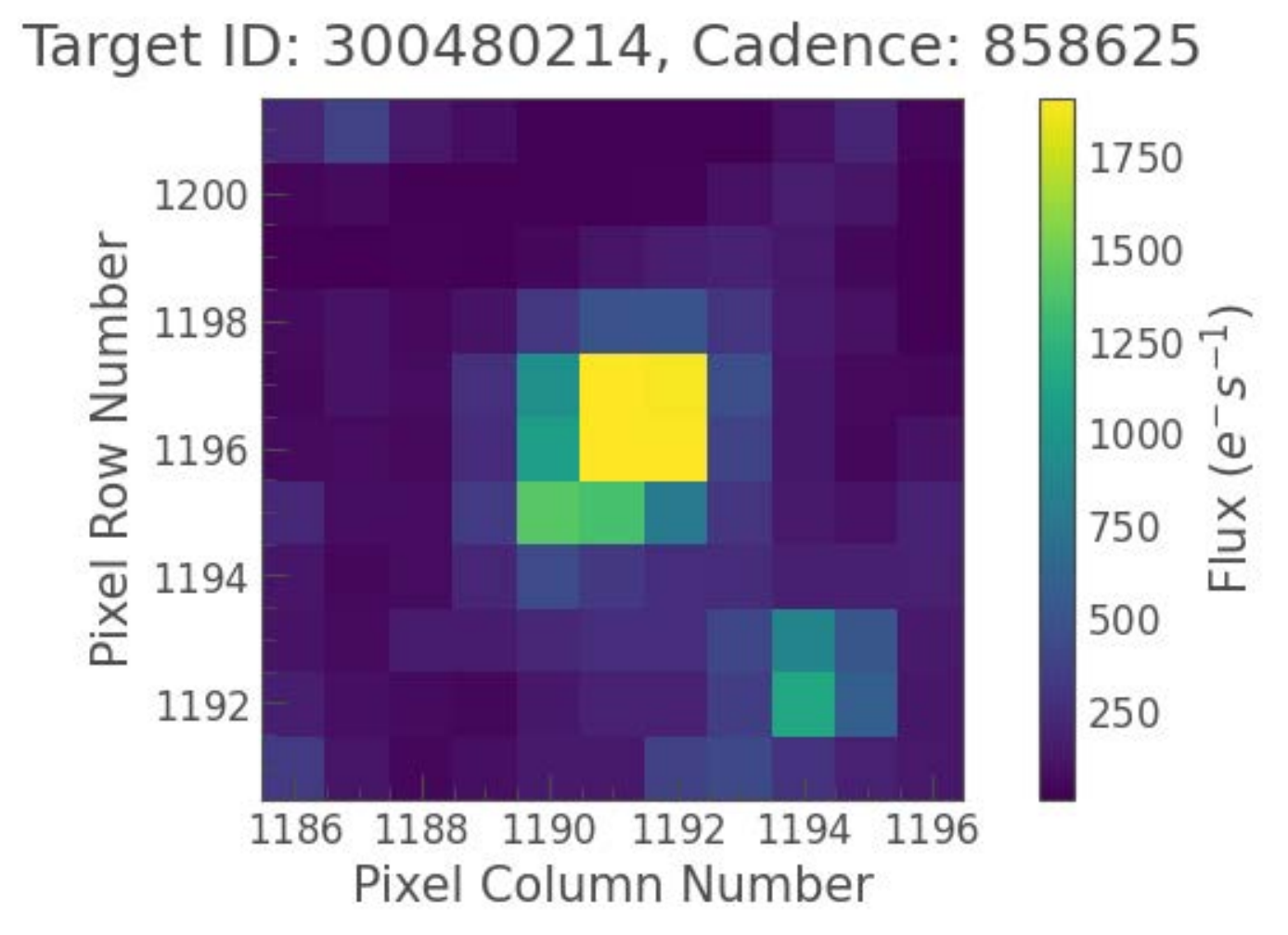}
\caption{TESS Sector 41 pixels on the CCD camera obtained in Python with Lightkurve \citep{2018ascl.soft12013L} and Matplotlib \citep{2007CSE.....9...90H} packages.}
\label{fig:Plot41}
\end{minipage}
\ \hspace{0mm} \hspace{1mm} \
\begin{minipage}[b]{7.5cm}
\centering
\includegraphics[width=7.5cm]{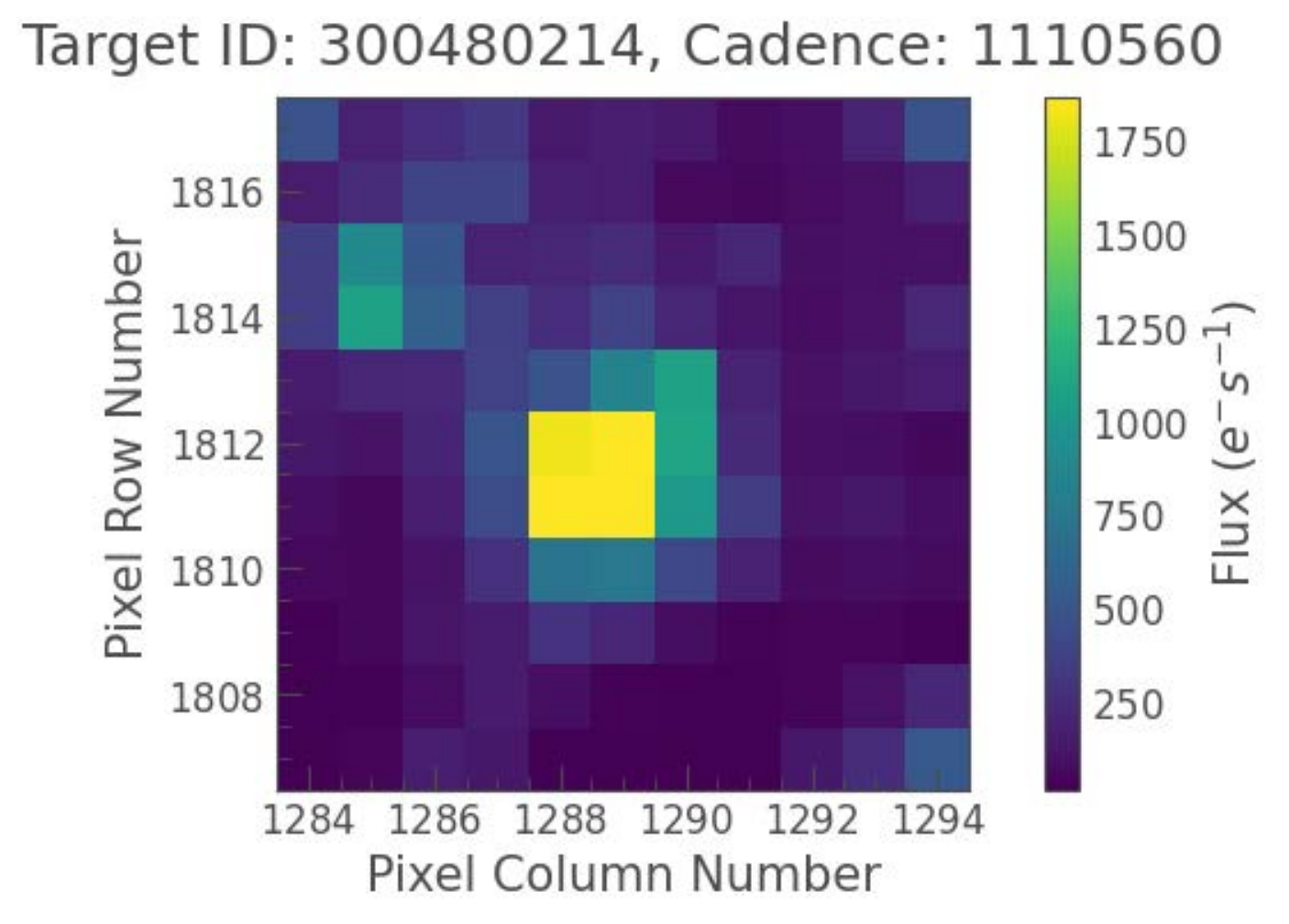}
\caption{TESS Sector 54 pixels on the CCD camera obtained in Python with Lightkurve \citep{2018ascl.soft12013L} and Matplotlib \citep{2007CSE.....9...90H} packages.}
\label{fig:Plot54}
\end{minipage}
\end{figure}

Using the MAST portal \citep{2011AAS...21734407C}, at coordinates RA(J2000) $19 : 52 : 19.13$ DEC(J2000) $+23 : 29 : 59.7$, 
raw data was obtained for star identified as TIC 300480214. 
In addition, from the TESS observation FITS, the CCD camera pixels with which \textit{MaGiV-1} was observed were obtained, and the color indicates the amount of flux in each pixel\footnote{Electrons per second}, as shown in Figure \ref{fig:Plot41} and Figure \ref{fig:Plot54} \citep{RickerGR(2014)}.\\

\vspace{0.5cm}

Before the light curve analysis, the offset 2457000 was added to the BJD values from the fits files. \citep{2010PASP..122..935E}.

In addition, brightness was converted from flux to magnitudes using the following law \citep{2000eaa..bookE1939B}:

\begin{equation}
Mag = -2.5log(Flux)
\end{equation}

\hspace{2cm}

\begin{figure}[ht]
\centering
\includegraphics[width=14cm]{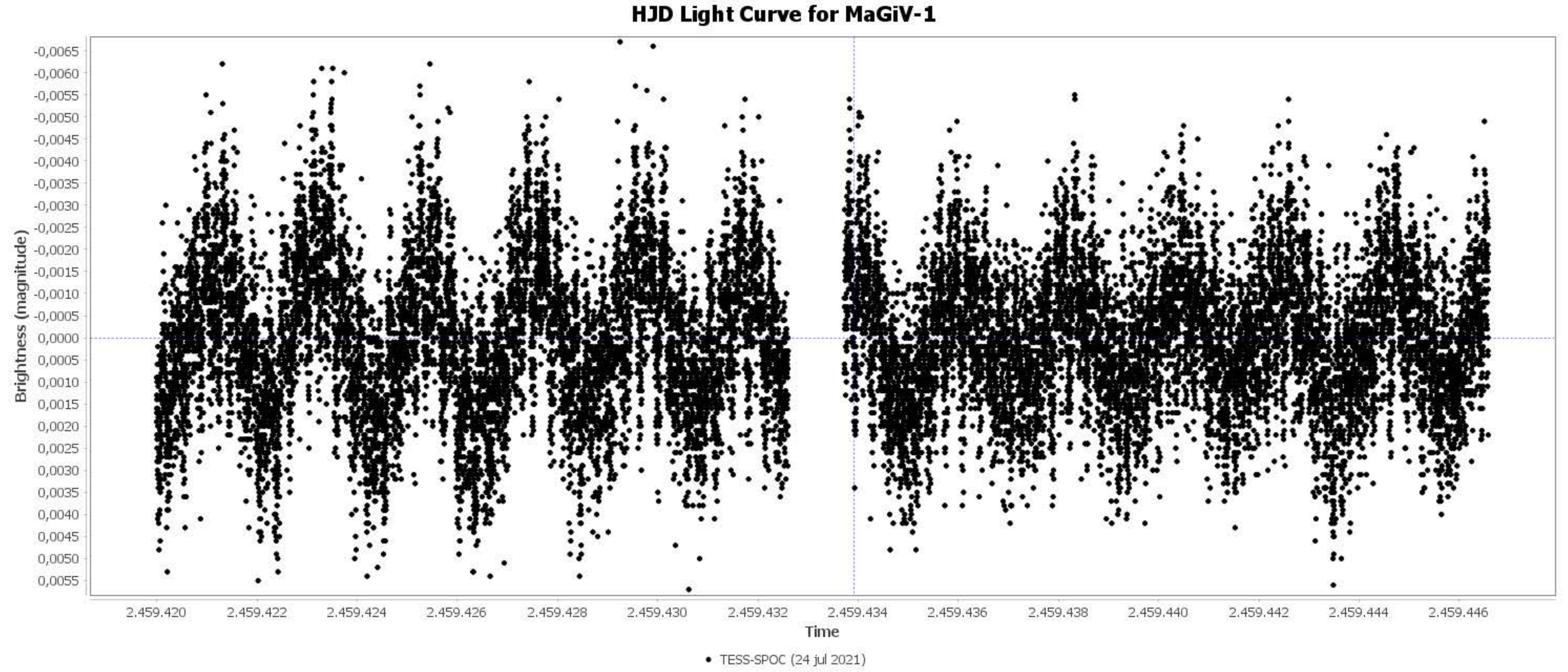}
\caption{Full TESS Plot for \textit{MaGiV-1} (Observation on July 24, 2021 in HJD time domain). Plot obtained with \textit{VStar} software \citep{Benn(2012)}.}
\label{fig:1stJD}
\end{figure}

\begin{figure}[ht]
\centering
\includegraphics[width=14cm]{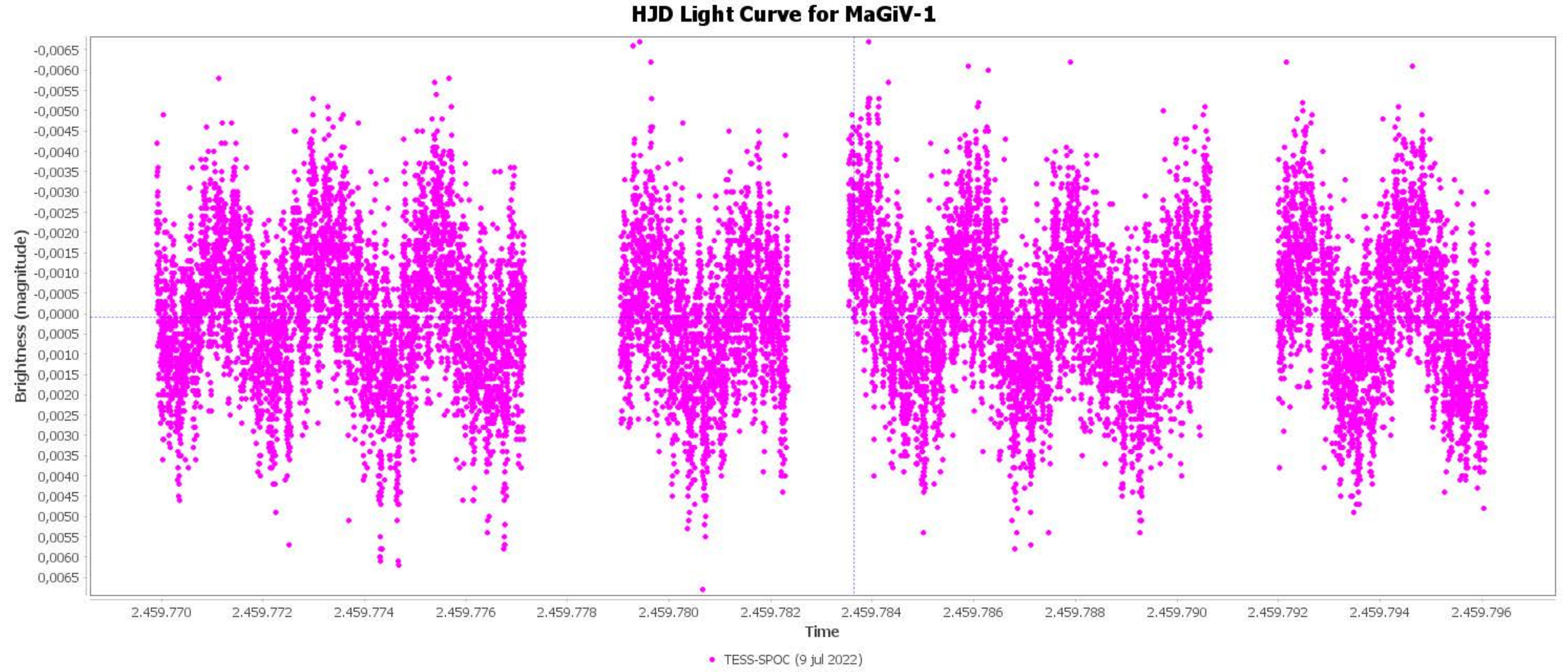}
\caption{Full TESS Plot for \textit{MaGiV-1} (Observation on July 9, 2022 in HJD time domain). Plot obtained with \textit{VStar} software \citep{Benn(2012)}.}
\label{fig:2ndJD}
\end{figure}

Finally, a normalized magnitude around zero was used, obtained by subtracting each measure from the average of the entire data set, highlighting the very small brightness variation. It can be shown in Figure \ref{fig:1stJD} and Figure \ref{fig:2ndJD} on 24 Jul 2021 and 9 Jul 2022 observations data respectively.

\newpage

The data, thus organized, were processed with Peranso software \citep{2016AN....337..239P}, obtaining the identification of two (or more) significant periods. The data, thus organized, were processed with the method ANOVA of the Peranso software \citep{1996ApJ...460L.107S}.

\begin{figure}[ht]
\centering
\includegraphics[width=14cm]{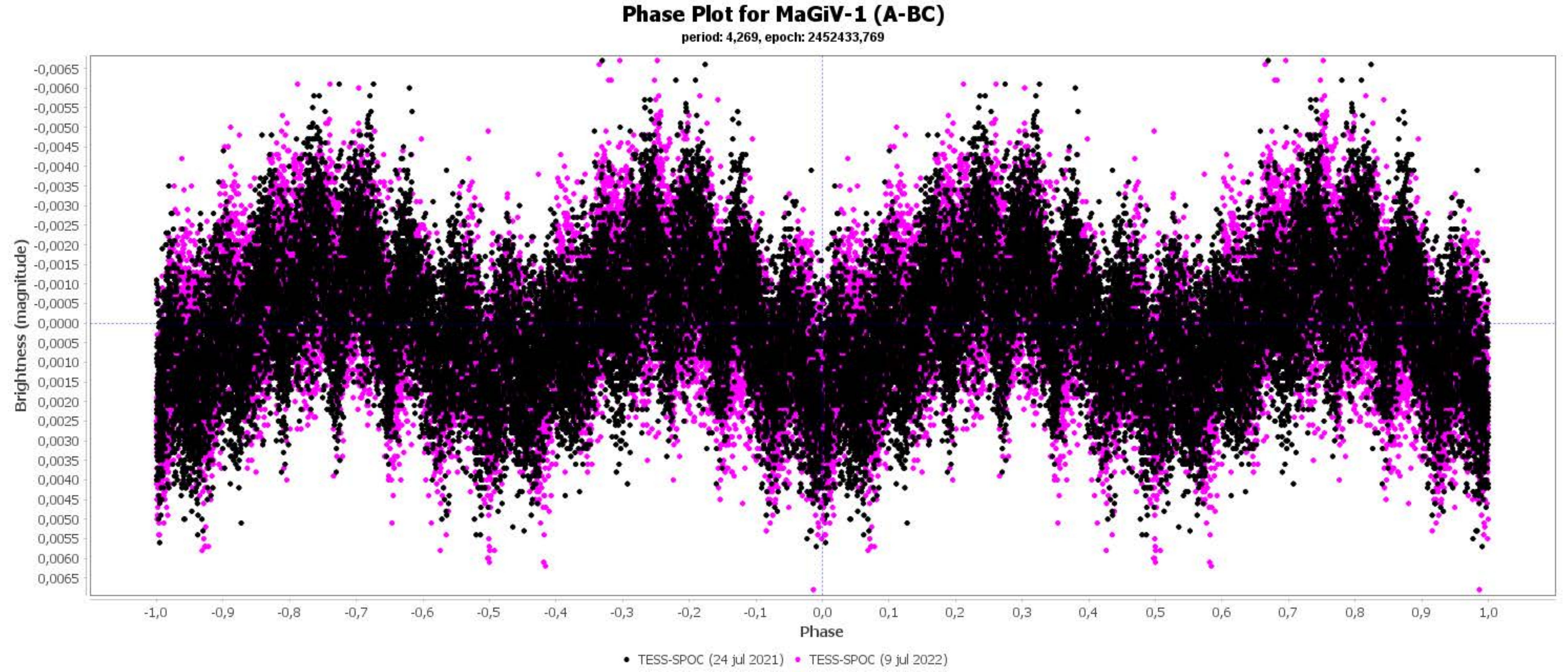}
\caption{Phase Plot for \textit{MaGiV-1} (1st ELL: Primary star ‘A’ with second couple consisting of ‘B’ and ‘C’ components). Plot obtained with \textit{VStar} software \citep{Benn(2012)}.}
\label{fig:1stELL}
\end{figure}

\begin{figure}[ht]
\centering
\includegraphics[width=14cm]{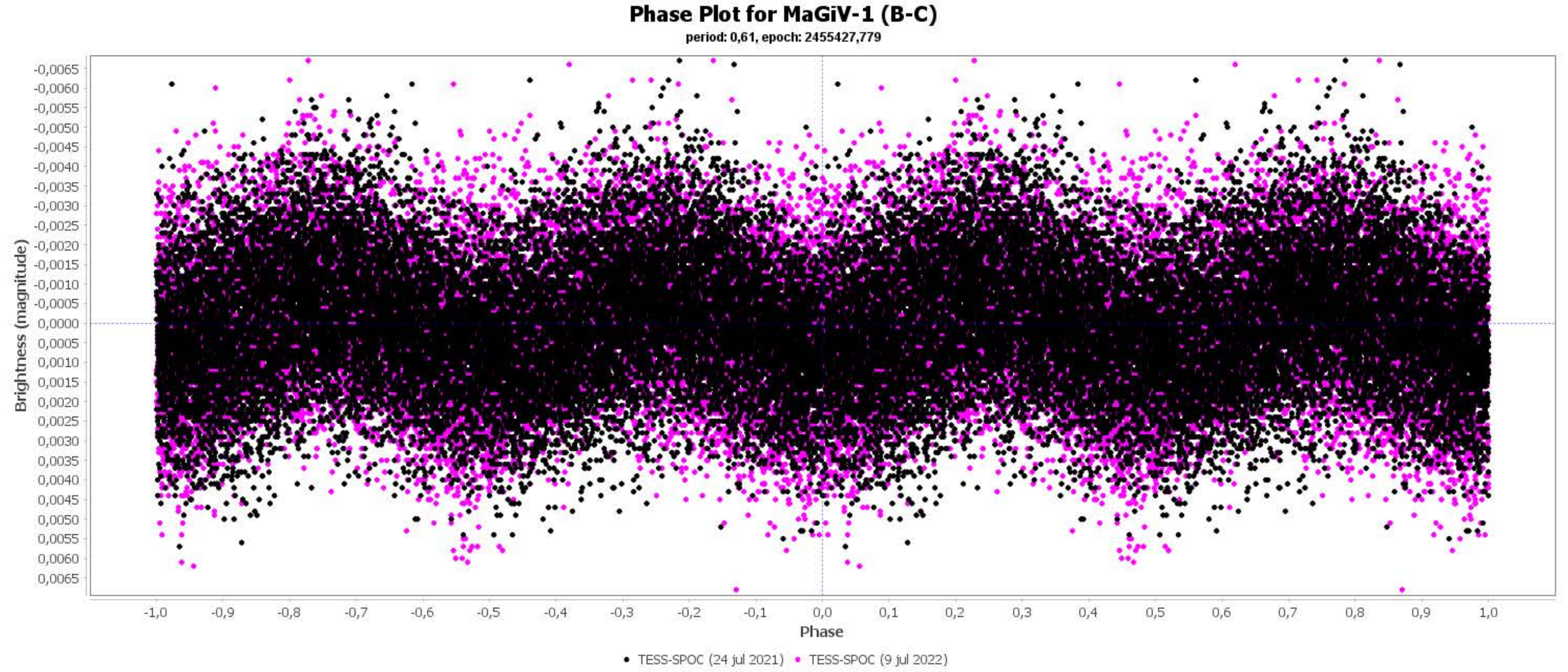}
\caption{Phase Plot for \textit{MaGiV-1} (2nd ELL: Couple consisting of ‘B’ and ‘C’ components). Plot obtained with \textit{VStar} software \citep{Benn(2012)}.}
\label{fig:2ndELL}
\end{figure}

Two signific periods were obtained to describe and characterize \textit{MaGiV-1} candidate triple-star system in the phase domain on two observation dates (24 Jul 2021 and 9 Jul 2022). First light curve in Figure \ref{fig:1stELL} characterized by a period $P_{A-BC} = (4.269 \pm 0.213)d$, shows two minima within an entire cycle (0 to 1) with a typical pattern of rotating ellipsoidal binary systems (ELL), but including obvious and repeated oscillations along the entire plot, with 0.004 delta magnitudes. These oscillations make sense when analyzing the second light curve in Figure \ref{fig:2ndELL} characterized by a period $P_{BC} = (0.610  \pm 0.031)d$, showing a common plot of ELL star systems with two minima within an entire cycle (0 to 1) and 0.003 delta magnitudes.


\vspace{1cm}

Since the depths of the two minima barely differ, the true period could also be half of $P_{BC}$ value, then $0.305d$. 

\newpage

Analyzing the periodogram, four significant values corresponding to the fundamental peaks were evaluated.

\begin{figure}[ht]
\centering
\includegraphics[width=14cm]{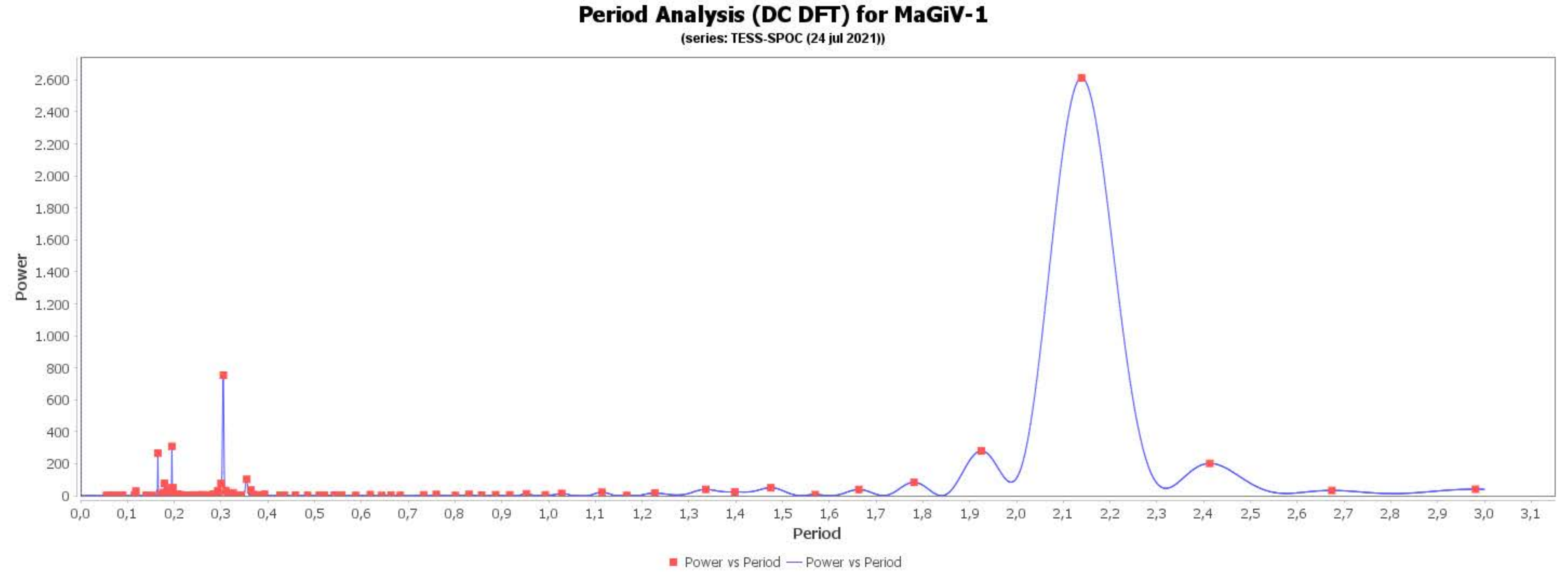}
\caption{Periodogram for \textit{MaGiV-1}. Plot obtained with \textit{VStar} software \citep{Benn(2012)}.}
\label{fig:pergram}
\end{figure}

The periodogram first shows two significant periods that could be interpreted as half of the orbital periods of the A-BC pair $P_{A-BC}=(4.269 \pm 0.213)d$ and the BC pair $P_{BC}=(0.610 \pm 0.031)d$. In addition, two periods below $<0.2d$ are present $P_{puls_{1}} = (0.195 \pm 0.050)d$ and $P_{puls_{2}} = (0.165 \pm 0.050)d$, which could represent \textit{Delta Scuti} type pulsations of one of the components. Since some Delta Scuti stars also show pulsations in the GDOR region \citep{2022A&A...666A.142S}, it cannot be completely ruled out that the presumed period of the BC pair is a pulsation frequency, even if this seems unlikely.

\section{Results}

\textit{MaGiV-1} variable star was identified through data-mining from the MAST website, where it was possible to view TESS photometric observations lasting 26.594 days, therefore reference series \textit{TESS\_TIC\_00000300480214\_Sector\_41\_PDCSAP} and \textit{TESS\_TIC\_00000300480214\_Sector\_54\_PDCSAP} were used.
It were extracted BJD (Barycentric Julian Date) time domain and the \textit{detrend} flow PDCSAP\_FLUX \citep{2016ComAC...3....6T}.
In Download Data Products (DDP) only infrared photometry (Ic) was extracted, using FITS Viewer software, a graphical program for viewing and editing any FITS format image or table \citep{WP(2020)}, and choosing the TIME and PDCSAP\_FLUX columns \citep{2011AAS...21734407C}.
The data thus obtained were appropriately converted and processed using Peranso software \citep{2016AN....337..239P} and analysing it with Fourier Transform derived by ANOVA method, using periodic orthogonal polynomials to fit the observations and statistics of variance analysis, defining their quality. This greatly improved the sensitivity of spike detection by dampening aliasing periods \citep{1996ApJ...460L.107S}.
It was identified a first period $P_{A-BC} = (4.269 \pm 0.213) d$ with 0.007 mag amplitude and a second period $P_{BC} = (0.610 \pm 0.031) d$ with 0.003 mag amplitude. From obtained light curves and about variable stars guidelines classification in Variable Stars Index (VSX) \citep{2020JAVSO..48..102O}, it was possible to classify the \textit{MaGiV-1} as a candidate triple star system ELL+ELL type candidate, then a triple-star system with ellipsoidal components whose luminosity variation changes with the orbital period.

Evaluating corresponding delta-magnitude for the light curves of ELL pairs respectively, it can be used the Period-Luminosity (P-L) relationship in low amplitude conditions \citep{2014AcA....64..293P}. This consideration makes it possible to define the overall brightness of two binary systems as a whole through a linear extrapolation that takes into account the orbital period.
Using the relationship

\begin{equation}
    W = -2.78(\pm 0.08)logP+19.30(\pm 0.18)
    \label{w}
\end{equation}

it was obtained following results

\begin{equation}
    W_{bc} = (19.90 \pm 0.11)mag
    \label{wbc}
\end{equation}

\begin{equation}
    W_{a-bc} = (17.55 \pm 0.16)mag
    \label{wabc}
\end{equation}

Where $W_{bc}$ is the overall brightness of the inner binary system and $W_{a-bc}$ is the overall brightness of the outer binary system. With these results, it was defined the ‘\textit{giant-giant}’ type components, in according by linear extrapolation model for low-amplitude proposed.


Finally, \textit{MaGiV-1} distance was obtained through the relationship linking it with the magnitudes resulting from the photometric analysis\\
\begin{equation}
    d = 10 ^ {(m_v-M_v + 5-A_v) / 5}
    \label{distanza}
\end{equation}

where $m_{v}$ and $M_{v}$ are relative and absolute magnitudes respectively.

The information provided by GAIA regarding parallax was used $ p = 0.835079181 mas $ and some features related to field such as extinction $ Av = (1.40 \pm 0.56) mag $ and magnitude of source $ m = (10.144 \pm 0.557) mag $, it was obtained \citep{2006JAHH....9..173H}.

Distance evaluated is $d = (629 \pm 245) pc$ and comparing it to the GAIA distance $d_ {GAIA} = 1018 pc$, shows an underestimation, because GAIA distance derived from parallax,  brings with it a more significant detection uncertainty, in fact, the index $RUWE = 2.387$ \citep{2020yCat.1350....0G} is much more than 1 (value of high affidability).


\end{sloppypar}

\end{document}